\documentstyle[12pt]{article}
\def\x{{\bf x}}
\def\v{{\bf v}}
\def\g{{\bf g}}
\def\a{{\bf a}}
\begin{document}
\begin{flushright}
titcmt-94-4 (chao-dyn/9405003)
\end{flushright}
\begin{center}
Numerical Study of Granular Turbulence  and\\
the appearance of $k^{-5/3}$ energy spectrum without flow\\
\vspace{1cm}
Y-h. Taguchi\\
\vspace{1cm}
Department of Physics,
Tokyo Institute of Technology\\
Oh-okayama, Meguro-ku, Tokyo 152
Japan\\
\end{center}

\vspace{1cm}

\noindent{\bf Abstract}

\noindent The vibrated bed of powder,
 a  vessel that mono disperse
glass beads fill and a loud speaker shakes,
is investigated numerically
with the distinct element method,
a kind of molecular dynamics.
When the bed is heavily shaken,
the displacement vectors of powder
have the power spectrum
with the dependence upon
the wave number $k$ as $k^{-5/3}$.
The origin of this spectrum
is suggested to be
the balance between 
the injected and dissipative energy,
analogous to the proposal by Kolmogorov
to explain the $k^{-5/3}$ energy
spectrum observed in the fluid turbulence.
Furthermore,
the same spectrum still appears
even without flows of powder.
Thus Kolmogorov's argument
is more universal than believed before.

\vspace{1cm}

\pagebreak

\section{Introduction}

For these a few decades,
the non-linear physics,
which has remained untouched over
several years since
there is not any
convenient method to
investigate it, has
become one of the central
topics in the modern physics.

This is because
the phenomenological approach provided
by the statistical physics
allows us to investigate them\cite{Kuramoto,CML}.
The basic philosophy underlying this approach
is:
'Some phenomena are independent of
the details of the specific systems;
Hence, the phenomenological
approach will be valid.'

For example,
the amplitude equation and
the phase dynamics approach\cite{Kuramoto}
can describe many
phenomena ranging from
the fluid dynamics to the chemical reaction.
However, the validity of these approaches
is restricted to the
weak non-linear regime
where only a few degrees of freedom
survive.

On the other hand,
the systems where many degrees of freedom
still survive and couples with each 
other---the subjects of the most 
interest among non-linear physicists---are attacked by the
phenomenological numerical approaches, 
like the coupled map lattice\cite{CML}.
Although they
exhibit many interesting
phenomena,
they lack the direct
 relations to the real systems.
At the moment, there is not any
general tool to study the real non-linear systems
with many degrees of freedom surviving.

Instead of trying to find general tools,
one can  specify phenomena where many degrees of freedom survive,
which appears in a real system 
with the easily treated model.
It will become a start point to construct the general theory.

An example among the phenomena that have the surviving many degrees
of freedom
is
the $k^{-5/3}$ energy 
spectrum in the fluid turbulence,
which is first proposed by Kolmogorov\cite{Monin}.
In his theory, he has predicted that
the energy spectrum of the fluid turbulence,
in the high enough wave number regime,
has the dependence upon the wave number $k$
as $k^{-5/3}$, 
which
was confirmed later experimentally and
numerically.
His theory
is valid for almost all
kinds of fluid turbulence:the experiments in the laboratories,
the atmospheric flows---wind,
and the flows in the tide.
His success strongly supports
the belief mentioned above:the details are not important.
However, his theory is too general
to make clear the mechanism
from the dynamical points of view;it does not have any direct
relationship
to the
Navier-Stokes equation:the basic equation of fluid motion.

The attempts to solidify the foundation of
his theory have not yet succeeded,
since it is hard to
obtain enough informations
of the local structures of the fluid turbulence.
Experimentally,
the simultaneous measurements of 
the velocity over the whole region
is impossible.
Only variables observed experimentally
is the time sequence of the velocity
at a specific point,
which is interpreted as the
spatial structure of the velocity
by assuming unjustified Teylor's
frozen turbulence hypothesis\cite{Monin}.
Numerically, although
the detail of the velocity field is
easily obtained,
the limited computational resources
makes the production of
fully developed fluid turbulence
difficult.

Even phenomenological approaches
are hardly applied to it.
The amplitude equation or
the phase dynamics approach
cannot be applied to $k^{-5/3}$
energy spectrum
where many degrees of freedom
is essential.
The phenomenological numerical
approach has succeeded
in reproducing
them\cite{Ohkitani,LVTM},
but could not explain the mechanism well
due to the lack of the direct relationship to
the experiment.
Hence,
for the fully understanding of the
Kolmogorov's theory, it is hopeful
to find the suitable substitution that
obeys Kolmogorov's law and
can be treated easily numerically and
experimentally.

In this paper,
the $k^{-5/3}$ energy spectrum
is investigated in such a substitution:the granular flow.
The granular systems,
which many physicists are recently interested 
in\cite{Science,review1,review2,review3},
are the systems whose
number of degrees of freedom is
 limited
enough to be integrated by the direct numerical simulations.
For example, the number of degrees of
freedom, even in the real experiments, is $10^6$,
which should be compared with
the Avogadro number $10^{23}$
that represents those of
typical materials like fluids, liquids, and solids.
The numerical
simulations of the system with
one hundred degrees of freedom
reproduce the non-trivial
properties originating in the strong non-linearity
as seen in the next section.
Therefore, we can expect the experimental correspondence
between the numerical model and the real experiments.

The organization of this paper is
as follows.
In Sec.\ref{sec:bed},
the results obtained numerically for
the vibrated bed of powder
is summarized.
In Sec.\ref{sec:model},
to investigate the origin of the $k^{-5/3}$
law,
the new model that
 has
the experimental correspondence
is introduced.
Discussions and conclusions
will be presented in Sec.\ref{sec:dis}.

\section{Numerical Granular Turbulence in Vibrated Bed of Powder}
\label{sec:bed}

\subsection{The experiment of vibrated bed of powder}
The experimental setup
of vibrated bed of powder is
as follows\cite{Evesque,Laroche}.
The flat vessel with the
horizontal dimension of about $10$ cm
is filled with granular matter:typically, mono disperse glass
beads having the diameter of about less than
$1$ mm.
The bed is shaken vertically
with the acceleration comparable 
to the gravity.
The speaker is usually used for this purpose,
the frequency of the vibration
is about from  $10$ to $100$ Hz,
and the harmonic dependence upon time $t$
is assumed;that is,
the vertical displacement
of the bed is as $b \cos \omega_0 t$.
Although most experiments are performed
in the three dimensional setups,
the few did experiments in two dimensions.

When the acceleration amplitude, $\Gamma=b\omega_0^2$,
exceeds some critical value 
$\Gamma_c$ slightly larger than gravity acceleration $g$,
the bed exhibits several
non-trivial dynamical
instabilities:the heaping, the convection\cite{Evesque,Laroche,tag,Gallas}, 
the capillarity\cite{Akiyama}, 
the surface fluidization\cite{tag,Evesque.epl,Luding},
the Brazil nuts segregation\cite{Rosato,Duran,Jullien,Ohtsuki,Knight},
 and the standing waves\cite{Douady,Pak,Swinney,Hubler},
and so on.

\subsection{Numerical Method---Distinct Element Method}
The distinct element method (DEM),
or the particle element method, was named analogous to
the finite element method.
Unlike in the finite element method,
each element in DEM can move freely, 
but like in the finite element method
represents
the macroscopic property of the material.
The individual granular particle is treated as an element 
obeying the  visco-elastic
equation,
\begin{equation}
\ddot{\x_i} = - \sum_{j=1}^{N} \theta ( d-\mid \x_i -\x_j \mid )
\left [ k_0 \left \{ \x_i-\x_j -d 
\frac{\x_i-\x_j}{\mid \x_i -\x_j \mid} \right \}
+ \eta ( \v_i -\v_j ) \right ] - \g,
\label{eq:basic}
\end{equation}
where 
$N$ is the total number of
granular particles, $\x_i$ is the position vector
of the $i$th particle,
$\g$ is the gravity acceleration, and $\v$ is velocity.
 Each particle has a diameter of $d$.
Because of a step function $\theta(x)$, particles interact  only when
they contact  with each other, which represents
the discrete nature of the granular matter.
$k_0$ and $\eta$ are the elastic constant and the viscosity coefficient,
respectively.
This visco-elastic material constants give coefficient of restitution, $e$, 
and collision time $t_{col}$ when two particles collide head-on;
$e=\exp(-\eta\pi/\omega)$ and $t_{col}=\pi/\omega$ where
$\omega = (2k_0-\eta^2)^{1/2}$.
In this sense, the visco-elasticity is not the real material property, but
the phenomenological one.
The DEM can be also considered as a simple molecular dynamics simulation
having the above interactions.

Cundall and Struck\cite{Cundall},
to investigate geological properties, 
 has first proposed DEM 
which has come to have several versions later.
However, in this paper, this simplified version,
which can reproduce
the convection\cite{tag,Gallas}---the most non-trivial phenomenon in
the vibrated bed of powder,
is used to study the vibrated bed of powder.

The above equations are integrated by the
Euler scheme where the time increment
$\Delta t$ changes at each step so that
the displacement  during $\Delta t$ of each particle
does not exceeds some fixed value $\alpha$ (See appendix \ref{app:deltat}).

To simulate the vibrated bed of powder,
the granular particles move within the two dimensional 
space---to save the cpu time--with 
the bottom oscillating as a function of time $t$,
$b \cos \omega_0 t$.
In this representation the
acceleration amplitude $\Gamma$, the strength of 
the external driven force, is $b \omega_0^2$.
When  the bottom collides with a granular particle,
it reflects the particle with the elastic constant $k_0$
without the dissipation.

\subsection{Numerical results in the vibrated bed of powder}

First, the numerical behavior of the vibrated bed of powder
obtained in the previous publications\cite{tag,tag2,tag.epl} is summarized.
When $\Gamma$ exceeds the gravity acceleration,
the convection and the surface fluidization are observed\cite{tag}.
The fluidized region where
the flow exists has the finite depth 
which increases as $\Gamma$ increases and
coexists with the solid region where the powder does not flow
(See Fig.\ref{fig:fluidized}).
\begin{figure}
\caption{The process of fluidization.
When $\Gamma < \Gamma_c$,
the bed consists of the solid region.
As $\Gamma$ exceeds $\Gamma_c$,
the fluid region appears,
but the solid region still remains.
For larger $\Gamma$,
the bed is fully fluidized.}
\label{fig:fluidized}
\end{figure}
To understand the dynamics of the vibrated bed of powder,
 we consider the granular flow in detail.

The granular flow is defined as
the displacement of particle relative to each other.
Thus, the motions like
the translation and the expansion--not giving rise to the
exchanges among particles--are not regarded as the flow
in this subsection. 

The numerical setup to measure the flow is shown in
Fig.\ref{fig:setup},
while the two dimensional space is horizontally periodic
and has the oscillating bottom.
\begin{figure}
\caption{Numerical setups used in this section.
The bed is horizontally periodic with the period of $L_h$, and
the lid stays on the bed.
}
\label{fig:setup}
\end{figure}
On the granular layer lies the lid
which never tilts and does obey  the equation of motion
\begin{equation}
M\frac{d^2h}{dt^2}=-\sum_{i=1}^N \theta(y_i-h)k_0(h-y_i), 
\label{eq:lid}
\end{equation}
where $M$ is the mass of the lid,
$h$ and $y_i$ are the vertical coordinate (height) of the lid
and the $i$th particle, respectively.
Each particle collides with the lid elastically---with $k_0$---and
never pass through it.
The lid suppresses the surface diffusion which causes
the  much larger displacement than the granular flow does.

The lid is also necessary to know 
when to record the positions of particles to
calculate the granular flow.
When vibrating, the bed expands and contracts repeatedly and
thus does not keep the constant volume.
To remove the contribution of these motions to
the displacements, which are much larger than those of the granular flow, 
we define the measuring time $t_n$ when
the spacing $\Delta h(t)$  (See Fig.\ref{fig:setup})
between the lid and the bottom has
some definite value, $h_0$.
It enables us to measure the displacement
under the condition of the constant volume
and to exclude the contribution of the volume changes
to the displacement of each particle.
Thus the definition of the displacement vector
is as
\begin{equation}
\Delta \x_i^{(n)}\equiv\x_i(t_{n+1})-\x_i(t_n).
\end{equation}

The typical displacement vectors
in the fully fluidized bed with $\Gamma=2.19$
($\omega_0=2\pi/6, b=2.0, N=1024,
L_h=128, e=0.8, t_{col}=0.1, d=2.0, g=1.0, M=100, 
h_0=34.0$\footnote{For
the dense packing the particle forms the triangular lattice
which has the width of $L_h/d=64$ and
the hight of $N/(L_h d) =16$.
Thus the hight of the bed under the dense
packing is $\frac{\sqrt{3}}{2} \times d \times \{(N/L_h d)-1\}
\sim 26.0$.})
is shown in Fig.\ref{fig:flow}.
\begin{figure}
\caption{Typical displacement vectors of fully fluidized bed
($\Gamma=2.19$)}
\label{fig:flow}
\end{figure}
The spatial structure looks 
the velocity field in the turbulent fluid
having the power spectrum 
with the $k^{-5/3}$  dependence upon
the wave number $k$.

To compare it with the spatial structure of the
velocity field in the fluid turbulence,
Fourier power spectrum is calculated from the
displacement vectors.
First, I divide whole two dimensional space into
$d$ (horizontal) $\times$ $\sqrt{3}d/2 $(vertical) cells.
Each cell moves with the bottom and
having numbers $(X,Y)$ where $X,Y$ are positive integers.
The cell $(X,Y)$ covers the area $ d(X-1/2) < x < d(X+1/2),
d(Y-1/2) < y < d(Y+1/2) $, where $x$ and $y$ are horizontal and vertical
coordinates that have its origin on the bottom.

I define displacement vector on each cell as:
\begin{equation}
\Delta {\bf x}_{X,Y}^{(n)} \equiv \sum_{i \in (X,Y)}  \Delta {\bf x}_i^{(n)},
\end{equation}
where summation runs over 
only particles in the cell $(X,Y)$.
I calculate Fourier power 
spectrum for $\Delta {\bf x}_{X,Y}^{(n)}$  on each layer 
\begin{equation}
S(K,Y) \equiv \langle
\mid \sum_X \Delta x_{X,Y}^{(n)} \exp (-j 2\pi XK/L) \mid^2 \rangle_n,
\end{equation}
where $K$ and $L$ are integers,
$L$ is $L_h/d$.
$\Delta x_{X,Y}^{(n)}$ is the horizontal component of
$\Delta {\bf x}_{X,Y}^{(n)}$.
$j$ is a pure imaginary number.
The average $\langle \cdots \rangle_n$
runs over  30 periods.

Figure \ref{fig:spec} shows the dependence of power spectrum
upon the wave number $K$ and the height from the bottom, $Y$.
\begin{figure}
\caption{The dependence of the power spectrum   $S(K,Y)$ 
of the fully fluidized bed
upon the
wave number $K$ and the height from the bottom $Y$
($\Gamma =2.19$).
(a) $Y=1 \sim 5$,(b) $Y=6 \sim 10$,
(c) $Y=11 \sim 15$,(d) $Y=16 \sim 20$.
The straight lines indicate the $K^{-5/3}$ dependence.
For large $Y$, i.e. near the surface,
the spectrum becomes flat in the high wave number region.}
\label{fig:spec}
\end{figure}
The power spectrum near the surface
deviates from the straight line
and has the flat spectrum---the 
white noise---in the higher wave number region,
but for small $Y$, the power spectrum has the
power dependence upon $K$.
The slope of log-log plot is very close to -5/3
which coincides with that of the Kolmogorov's scaling theory.
Therefore, we can conclude that
the spatial structure in the vibrated bed is similar to that
in the fluid turbulence.

The origin of this turbulent
 motion in the fluidized region should exist
in the motion of solid region,
because as shown in Fig.\ref{fig:fluidized}
the solid region 
becomes fluidized gradually as $\Gamma$ increases.
To compare the motion in the solid region
with that in the fluidized region,
the power spectrum averaged over $64$ periods in the solid region
when $\Gamma=1.10$($b=1.0, h_0=30.0$,  and the remaining parameters
are identical to the above)
is shown in Fig.\ref{fig:spec2}.
\begin{figure}
\caption{The dependence of the power spectrum $S(K,Y)$
of the almost solidized bed upon the
wave number $K$ and the height from the bottom $Y$
($\Gamma =1.10$).
(a) $Y=1 \sim 5$.
(b) $Y=6 \sim 10$.
(c) $Y=11 \sim 15$.
(d) $Y=16 \sim 18$.
For details, see Fig.\protect\ref{fig:spec}.}
\label{fig:spec2}
\end{figure}
As seen in the typical displacement 
vector
 (Fig.\ref{fig:flow2}),
\begin{figure}
\caption{Typical displacement vectors of almost solidized bed.
($\Gamma=1.10$)}
\label{fig:flow2}
\end{figure}
almost all region is solid---no exchange between particles,
but the global structure of the power spectrum
is qualitatively similar;the power spectrum in the lower layers
has the power dependence upon $K$ with the
exponent $-5/3$, and
the white noise appears in the higher wave number region
for the upper granular layers.

In the following, I regard
this $k^{-5/3}$ spectrum
without flow as the origin of
that seen in the fluidized region
and make clear how the $k^{-5/3}$ spectrum
appears without the flow.

\section{The $k^{-5/3}$ power spectrum without flow}
\label{sec:model}

To understand the mechanism of the $k^{-5/3}$ power spectrum
without flow, a model in which the powder does not flow is proposed. 
Each element, the granular particle, 
forms a triangular lattice\footnote{Hence no granular flow---no
exchange between particles---is allowed at all}
as shown in Fig.\ref{fig:setup}
 and
interact,  visco-elastically 
as described in Eq.(\ref{eq:basic}) but without gravity $\g$,
with the six neighbors.
The bottom oscillates as before, but
the lid does not obey the equation of motion, Eq.(\ref{eq:lid}).
Instead, the lid oscillate as a function of time, $b_0 \cos \omega_0 t$.
That is, when the vertical coordinate of the bottom oscillate as
$b \cos \omega_0 t$, then the height of the lid $h=h_0+b_0 \cos \omega_0 t$.
Thus, $h_0$ becomes the average distance between the bottom and the lid.
At the initial stage, the granular particle lies
on the triangular lattice between the bottom and the lid
and moves obeying the Eq.(\ref{eq:basic}) with $\g=0$.

This numerical setup corresponds to
the horizontally vibrated bed in the real experiments.
One prepares the flat cell filled with the
glass beads and makes two side walls oscillate harmonically.
Hence, the following results are expected to be reproduced
in the experiments.

First, both the lid and the bottom oscillate with the same harmonic 
form, i.e. $b=b_0=1.0,\g=0.0$.
The $h_0$ is taken to be equivalent to the width
of the triangular lattice having $N/L$ 
rows, $\sqrt{3}/2 \times [(N/L)-1]$,
and thus the granular particle is packed without space between each other.
The remaining parameters are the same as used in the previous section.

Figure \ref{fig:spec.lt} shows the
time development of the energy spectrum
\begin{equation}
E(K,t) \equiv \langle
\mid \sum_{i=1}^L \Delta v_x(t;i,j) \exp (-j 2\pi iK/L) 
\mid^2 \rangle_{j,initial},
\end{equation}
where $v_x(t;i,j)$ is the $x$ component---the direction perpendicular to
the direction of the vibration---of the velocity at
the site $(i,j)$ at time $t$.
The average is taken over the layers---$y$ direction---and
the ten initial realization of the velocity
with the white noise energy spectrum.
One should notice that
it is the true energy spectrum---not
the power spectrum of the displacement vector
as in the previous section---thus
we can compare it with the Kolmogorov's argument.
\begin{figure}
\caption{The time development of the energy spectrum,
$E(K)$.(a) The initial state($t=1.5$) and
the early stage.(b) The medium stage where
the high wave number components start to
have flat part.
(c)The late stage.
The energy spectrum shows the system
reach the thermal equilibrium state.}
\label{fig:spec.lt}
\end{figure}
At the initial time $t=1.5$,
each granular particle has the small random velocities
and thus has the flat energy spectrum---the white noise.
As time proceeds,
the energy spectrum starts to incline
and comes to have power dependence upon
the wave number $k$.
Furthermore, 
its slope is very close to the -5/3.
However, for the medium stage,
the flat spectrum appears in the higher wave number region and
reaches the lower wave number region.
Finally in the later stage, the total energy spectrum has become flat.

This process can be interpreted as follows.
First, the energy starts to dissipate 
in the high wave number region where
the energy dissipation rate takes the large value.
It enables the system to reach
the stationary state
described by the Kolmogorov theory.
However, the further injection of the energy
dominates the dissipation
and the system reaches the thermal equilibrium
state:each mode has the same amount of the kinetic energy.
This is also seen in the
numerical simulation of the Navier-Stokes turbulence
when the dissipation is not large enough.
Therefore, the dynamics
in the vibrated bed of powder is
essentially equal to that of the fluid turbulence.

However, the $k^{-5/3}$ spectrum
appears only temporally,
contrast to the permanent appearance of it
in the previous section.
The dissipation in the vertically vibrated bed
may be larger because 
the bed is compressed between
the lid and the bottom.
To take into account this effect,
the value of $b$ is taken to be 0
while $b_0$ remains unchanged so
that the volume of the lattice changes.
Starting from the initial white noise,
the power spectrum  becomes $k^{-5/3}$ one
and returns to the flat one.
However, this time the power
spectrum recovers $k^{-5/3}$ dependence
from the white noise
and this process is repeated once a period
of the oscillation(Fig.\ref{fig:spec.lt2}).
Here the energy spectrum is averaged over the ten periods
to show the dependence upon the fraction of cycle.
\begin{figure}
\caption{The dependence of the energy spectrum $E(K)$
upon
the  fraction of
the cycle, $\omega_0 t= n\pi/6.(n=1,2,\ldots,6)$.
The $k^{-5/3}$ dependence is shown
for the comparison.}
\label{fig:spec.lt2}
\end{figure}
Thus $k^{-5/3}$ power spectrum can appear repeatedly if the
effect of the compression is considered.

\section{Discussion and Summary}
\label{sec:dis}

In the previous section,
the powder bed has the energy spectrum
with $k^{-5/3}$.
The results also suggest that the
dissipation plays the important role.
The behavior of the energy spectrum
is coincident with that
in the Navier-Stokes numerical simulation
when the dissipation is not large enough.
However, there is not the flow at all,
thus one cannot expect the
powder obeys the Navier-Stokes equation.
So, what is the relation to the Kolmogorov theory
proposed for the fluid flow?

Here the Kolmogorov's argument should be
reconsidered.
The essential assumption in the Kolmogorov's theory
is that  the balance between the energy input and the energy dissipation
governs the dynamics of system.
This means, the amount of total energy is as large as the product of
the energy dissipation rate---or the energy input rate---and the
characteristic time length.
Figure \ref{fig:ballance} shows 
 the total energy, the energy input rate,
 and the energy dissipation rate in the fluidized region
investigated in Sec.\ref{sec:bed}
as a function of time $t$.
\begin{figure}
\caption{
 the total energy, the energy input rate,
 and the energy dissipation rate
as a function of time $t$
(See appendix \protect\ref{app:caleng}).}
\label{fig:ballance}
\end{figure}
The total energy consists of 
both the elastic energy---among particles,
the bottom, and the lid---and 
the kinetic and gravitational energy
of particles and the lid.
The dissipation energy comes from the viscosity
between particles, and
the work done by the vibrating bottom
is considered as the energy input.
The product between the energy input rate
($\sim$ a few thousands)   and the period of the vibration (=6)
is about $10^4$. 
On the other hand, in Fig.\ref{fig:ballance},
the actual total energy is about $10^4$,
when we eliminate the gravitational
potential of the grand state($\sim 2 \times 10^4$).
Thus, the total energy included in the bed is almost equal 
to the energy injected each period.
This means, the balance between the energy input and the energy dissipation
dominates the dynamics of the system.
So far, the basic assumption
necessary for the Kolmogorov's theory
is satisfied.

The energy transport between the modes
with the different wave number
is also assumed to be caused by the non-linearity,
which the step function $\theta(x)$ in Eq.(\ref{eq:basic})
has.
Actually speaking,
Kolmogorov has never
used the existence of the flow
to derive his scaling form,
since it is apparent;the turbulence must flow.
Here, the steady state which he proposed in his
scaling theory can appear without the flow
because the vibrated bed of powder
can satisfy the basic requirements without the flow.

Another difference from the fluid is the dimensionality.
The $k^{-5/3}$ spectrum can appear only in the three dimension,
not in the two dimension where it appears in the powder.
However, again, Kolgomorov has not used
the dimensionality explicitly.
The main difference between
two and three dimension
is whether there is another conserved variable,
the enstrophy\footnote{The enstrophy is
defined as $\langle \omega(\x,t)^2 \rangle /2$,
where $\omega$ is the vorticity, $\hbox{rot} \v$.}.
The enstrophy is impossible to define in
the vibrated bed of powder
because in the discrete material
like the powder the spatial derivative,
which is necessary to calculate the enstrophy,
does not exists due to the lack of the continuity.
Hence the enstrophy, which does not exist,
cannot prevent the two dimensional system of the
powder from obeying Kolmogorov's argument.

Assuming the $k^{-5/3}$ power spectrum comes from
the Kolmogorov theory,
we can explain the power spectrum seen in Sec. \ref{sec:bed}.
For lower layers of the bed,
the compression is strong enough,
thus the clear $k^{-5/3}$ spectrum can be observed.
However, the upper layers are not compressed enough to
dissipate injected energy,
which makes the energy stay in the high wave number region
and causes the flat spectrum.
Figure \ref{fig:schem} shows the schematics of
the energy flow in the vibrated bed
considering the above discussions.
\begin{figure}
\caption{The schematics of
the energy flow in the vibrated bed.}
\label{fig:schem}
\end{figure}

In summary, the vibrated bed of powder
has the $k^{-5/3}$ power spectrum
of the displacement vector.
It comes from the local motions in the solid region
which the Kolmogorov theory can explain.
This means, I found that the steady state proposed by
the
Kolgomorov can appear even without flow;thus 
his theory is more universal than believed before.
The real experiments should confirm these results.

This finding will make the effort to understand
the $k^{-5/3}$ mechanism easier,
since the powder system does 
not flow---thus we can use lattice representation,
and the dimensionality is 
two---requiring smaller computational resources
than the three dimension does.
The efforts also may give rise to understand
the general feature of
the strongly non-linear system
where many degrees of freedom survives and
may give us the general method
to study such systems.

\section{Acknowledgement}
This study is supported by Hosokawa-Powder Technology Foundation,
 Foundation for Promotion of Industrial Science,
and Grant-in-Aid for Encouragement of Young Scientists
(05740252) from the
Ministry of Education, Science, and Culture, Japan.

\appendix
\section{Numerical method to integrate the equation of motion}
\label{app:deltat}
 
The method employed to integrate the equation of the motion
Eq.(\ref{eq:basic}) is Euler scheme,
\begin{eqnarray}
\x_i(t+\Delta t) & = & \x_i(t) + \v_i (t)\Delta t 
+ \frac{1}{2} \a_i (t) (\Delta t)^2\label{eq:Euler1}\\
\v_i(t+\Delta t) & = & \v_i (t) + \a_i(t) \Delta t,
\end{eqnarray}
where $\a_i$ is the acceleration of $i$th particle.
Eq.(\ref{eq:Euler1}) enables us to
relate the time increment $\Delta t$ to
$\Delta x_i(t) \equiv \mid x_i(t+\Delta t)-x_i(t) \mid$
and
$\Delta y_i(t) \equiv \mid y_i(t+\Delta t)-y_i(t) \mid$,
where $x_i$ and $y_i$ are the two components of $\x_i$ respectively.
Then
\begin{eqnarray}
\Delta t _{xi}&=&\frac{\sqrt{v^2_{xi}+2a_{xi}\Delta x_i}-v_{xi}}{a_{xi}}\\
\Delta t _{yi}&=&\frac{\sqrt{v^2_{yi}+2a_{yi}\Delta y_i}-v_{yi}}{a_{yi}},\\
\end{eqnarray}
where $v_{xi}$ and $v_{yi}$ are the $x$ and $y$ components of $\v_i$,
and  $a_{xi}$ and $a_{yi}$ are the $x$ and $y$ components of $\a_i$.
Using these equations, the representation of $\Delta t$ with $\Delta \x_i$,
$\Delta t$ takes the value of $\min_{\x_i} \Delta t_{\x_i}$
with the fixed $\Delta x_i$ and $\Delta y_i$. 
In the present simulation, $\Delta x_i = \Delta y_i = \alpha = 0.005d$.

\section{The calculation of the energy and dissipation}
\label{app:caleng}

In this appendix, the definition of several quantities
appearing in Fig.\ref{fig:ballance} is explained.

The total energy consists of the kinetic energy,
the elastic energy, and the gravitational potential energy.
The kinetic energy $E_{kinetic}$ is defined as,
\begin{equation}
E_{kinetic}=\frac{1}{2} M \v_{lid}^2 +\sum_{i=1}^N \frac{1}{2} \v_i^2, 
\end{equation}
where $\v_{lid}$ is the velocity of the lid.
The gravitational potential energy $U_g$ can be expressed easily, too,
\begin{equation}
U_g = Mgh + \sum_{i=1}^N gy_i.
\end{equation}
Here one should remember $h$ is 
the height---the vertical coordinate---of the lid.
The elastic energy $U_{el}$ consists of those among 
the particles, the bottom plate, and the lid,
\begin{eqnarray}
U_{el} & = &\frac{1}{2}k_0 
\left \{ 
\sum_{i,j}  \theta ( d-\mid \x_i -\x_j \mid )
( d-\mid \x_i -\x_j \mid )^2
  \right . \nonumber\\
& + & \sum_{i=1}^N \theta (b \cos \omega_0 t - y_i) 
(b \cos \omega_0 t - y_i)^2  \nonumber\\
& + & \left . \sum_{i=1}^N \theta (y_i -h) 
( y_i -h )^2
\right \},
\end{eqnarray}
Thus the total energy $E_{tot}$ can be expressed as
$E_{tot}=E_{kinetic}+U_g+U_{el}$.

The energy input rate $E_{input}$ is the amount of work
done by the bottom per unit time,
\begin{eqnarray}
E^{(t)}_{input} &= &\frac{1}{\Delta t}\sum_{i=1}^N  [
k_0 \left \{ \theta (b \cos \omega_0 t -  y_i(t) ) 
(b \cos \omega_0 t - y_i(t)) \right \}   \nonumber\\
&& \times b \{\cos \omega_0 (t+\Delta t) -\cos \omega t \}  ].
\end{eqnarray}
Typically, $\Delta t $ is taken to be $1 \times 10^{-5}$.
The energy dissipation $E_{dis}$ due to the viscosity
between particles $i$ and $j$ is 
$\eta \int \mid (\v_i-\v_j)\cdot d(\x_i-\x_j) \mid$
when two particles contact with each other.
Using $d(\x_i-\x_j)/dt = (\v_i-\v_j)$, we get 
$E_{dis}= \eta \int (\v_i-\v_j)^2 dt$.
In the present discretization
substituting $\v_i(t+\Delta t)=\v_i(t) +\a_i(t)\Delta t$,  
\begin{equation}
E_{dis}=\frac{\eta}{\Delta t} \sum_{i,j} 
\theta ( d-\mid \x_i -\x_j \mid ) \left \{ \Delta \v_{ij}^2 \Delta t +
\Delta \v_{ij} \cdot \a_{ij} (\Delta t)^2 +
\frac{1}{3} \a_{ij}^2 (\Delta t)^3 \right \},
\end{equation}
where $\Delta \v_{ij}=\v_i-\v_j$ and
$\Delta \a_{ij} =\a_i-\a_j$.

These above variables have the relation,
$E_{tot}(t+\Delta t)=E_{tot}(t)+\{E_{input}(t)-E_{dis}(t)\}\Delta t 
+ {\cal O}(\Delta t^2)$.


\begin{thebibliography}{99}
\bibitem{Kuramoto} Y. Kuramoto,
{\it Chemical Oscillations, Waves, and
Turbulence}(Springer, Heidelberg, 1984)
\bibitem{CML}K. Kaneko,
{\it Collapse of Tori and Genesis of Chaos in
Dissipative Systems}
(World Scientific, Singapore, 1986)
\bibitem{Monin}A.Monin and A.Yaglom,
Statistical Fluid Mechanics (MIT Press, Cambridge, 1971).
\bibitem{Ohkitani} M. Yamada and K. Ohkitani, Phys. Rev. Lett.
60 (1988) 983.
\bibitem{LVTM} Y-h. Taguchi and H. Takayasu, Physica D 69 (1993) 366.
\bibitem{Science} H. M. Jaeger and S. R. Nagel,
Science 255 (1992) 1523.
\bibitem{review1}Y-h. Taguchi, H. Hayakawa, S. Sasa, and H. Nishimori
eds., Dynamics of Powder Systems,
Int. J. Mod. Phys. B 7 Nos. 9 \& 10 (1993).
\bibitem{review2}D. Bideau and A. Hansen
eds., Disorder and Granular Media
(North-Holland, Amsterdam, 1993).
\bibitem{review3}A. Mehta ed., Granular Matter
(Springer, Berlin, 1993).
\bibitem{review4}C. Thornton ed., Powders and Grains '93
(A.A.Balkema Publishers, Rotterdam, 1993).
\bibitem{Evesque} P. Evesque and J. Rajchenbach,
Phys. Rev. Lett. 62 (1989) 44.
\bibitem{Laroche} C. Laroche, S. Douady, and S. Fauve,
J. Phys. (Paris) 50 (1989) 699.
\bibitem{tag} Y-h. Taguchi, Phys. Rev. Lett. 69 (1992) 1367.
\bibitem{Gallas} J. A. C. Gallas, H. J. Herrmann, and S. Soko\l owski,
Phys. Rev. Lett. 69 (1992) 1371.
\bibitem{Akiyama} T. Akiyama and T. Shimomura,
Powder Technology 66 (1991) 243.
\bibitem{Evesque.epl} P. Evesque, E. Szmatula, and J-P. Denis,
Europhys. Lett. 12 (1990) 623.
\bibitem{Luding} S. Luding, E. Clement,
A. Blumen, J. Rajchenbach, and J. Duran,
Phys. Rev. E 49 (1994) 1634.
\bibitem{Rosato} A. Rosato, K. J. Strandburg,
F. Prinz, and R. H. Swendsen,
Phys. Rev. Lett. 58 (1987) 1038.
\bibitem{Duran}J. Duran, J. Rajchenbach, and E. Cl\'ement,
Phys. Rev. Lett. 70 (1993) 2431.
\bibitem{Jullien}R. Jullien, P. Meakin, and A. Pavlovitch,
Phys. Rev. Lett. 69 (1992) 640.
\bibitem{Ohtsuki}T. Ohtsuki, Y. Takemoto, T. Hata, S. Kawai, and A. Hayashi,
Int. J. Mod. Phys. B 7 (1993) 1865.
\bibitem{Knight} J. B. Knight, H. M. Jaeger, and S. R. Nagel,
Phys. Rev. Lett. 70 (1993) 3728.
\bibitem{Douady} S. Douady, S. Fauve, and C. Laroche,
Europhys. Lett. 8 (1989) 621.
\bibitem{Pak} H. K. Pak and R. P. Behringer, Phys. Rev. Lett.
71 (1993) 1832.
\bibitem{Swinney} F. Melo, P. Umbanhowar, and H. L. Swinney,
Phys. Rev. Lett. 72 (1994) 162.
\bibitem{Hubler} Dinkelacker F.,H\"ubler A., and L\"uscher E.,
Biol. Cybern. 56 (1987) 51.
\bibitem{Cundall} P. A. Cundall and O. D. L. Strack, 
Geotechnique 29-1 (1979) 47.
\bibitem{tag2} Y-h. Taguchi, J. Phys. II (Paris) 2 (1992) 2103.
\bibitem{tag.epl} Y-h. Taguchi, Europhys. Lett. 24 (1993) 203.  
\end{thebibliography}
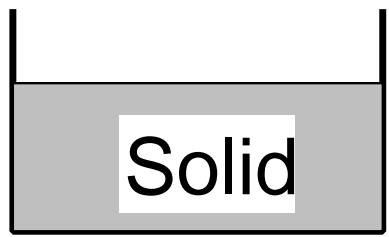
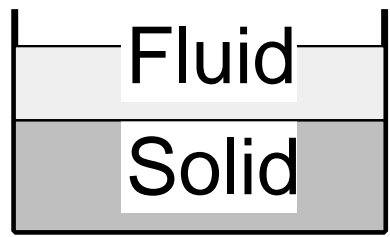
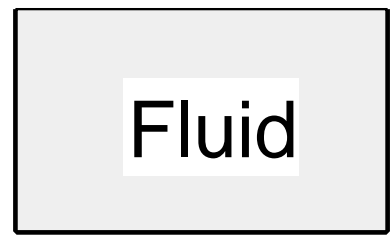
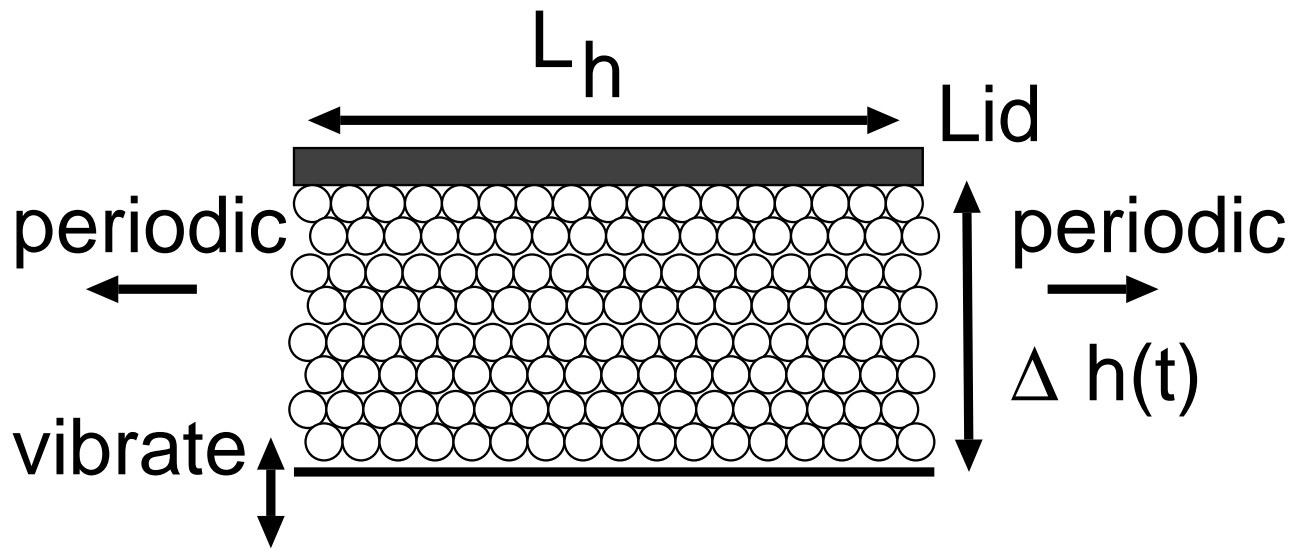
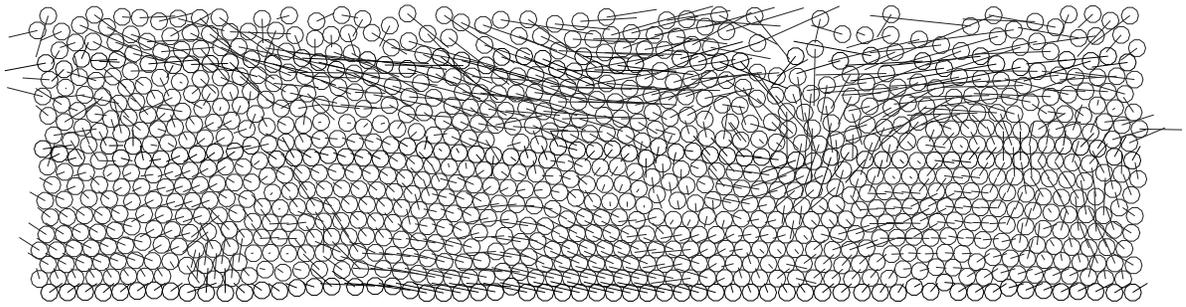
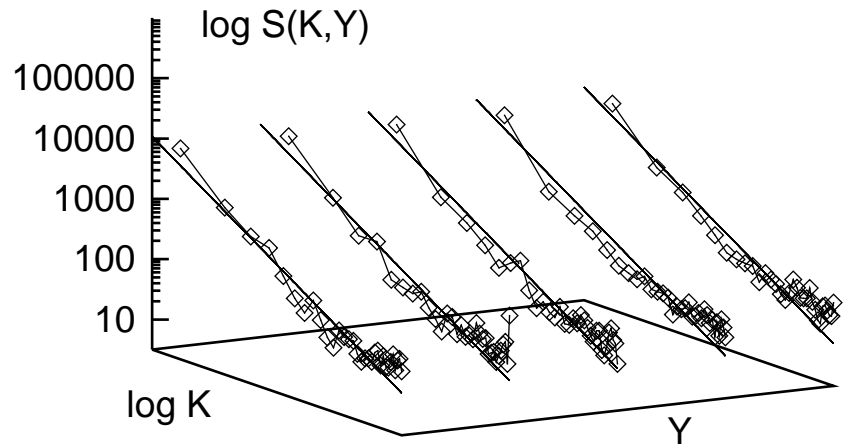
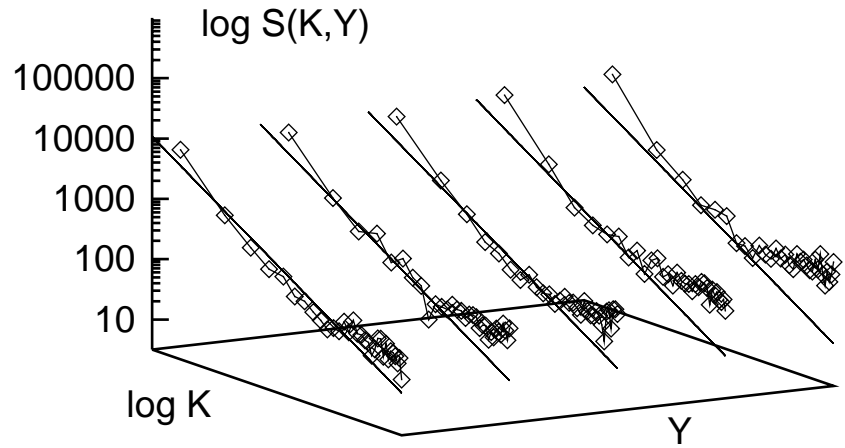
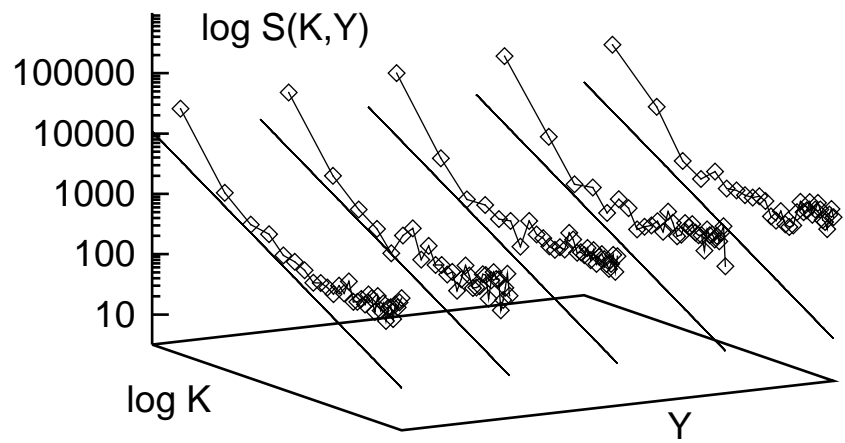
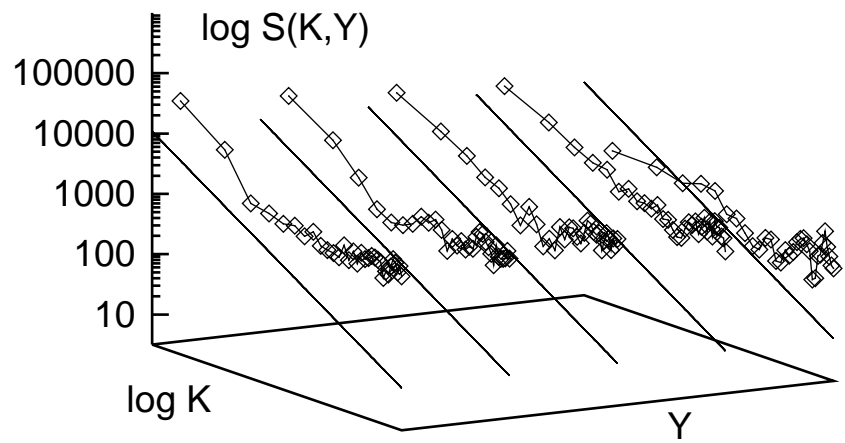
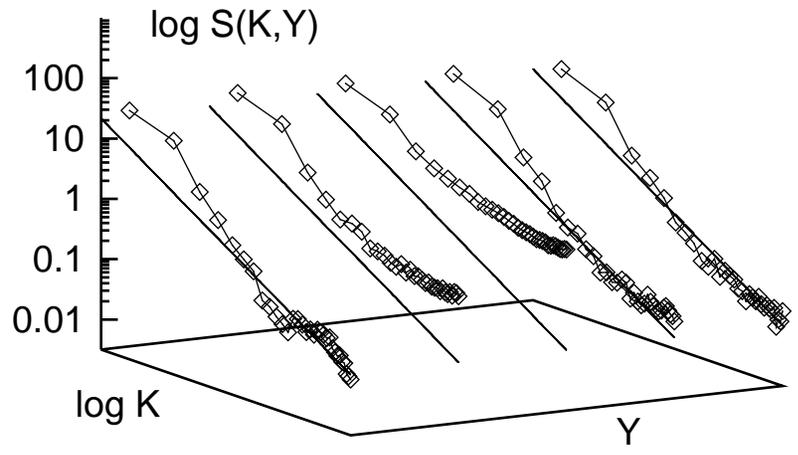
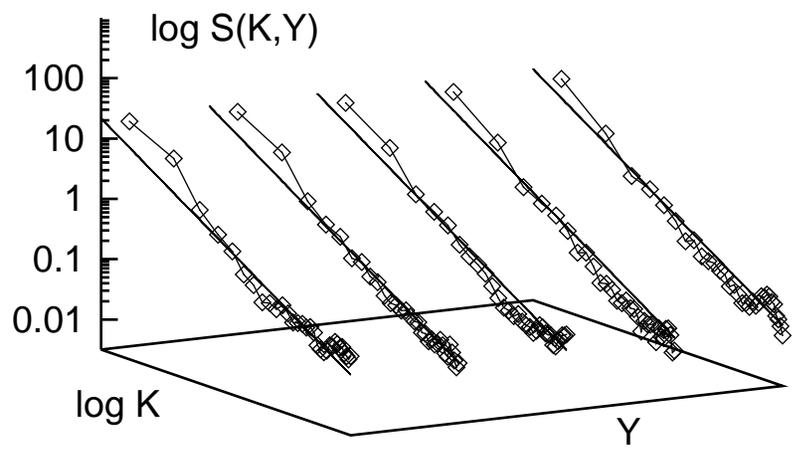
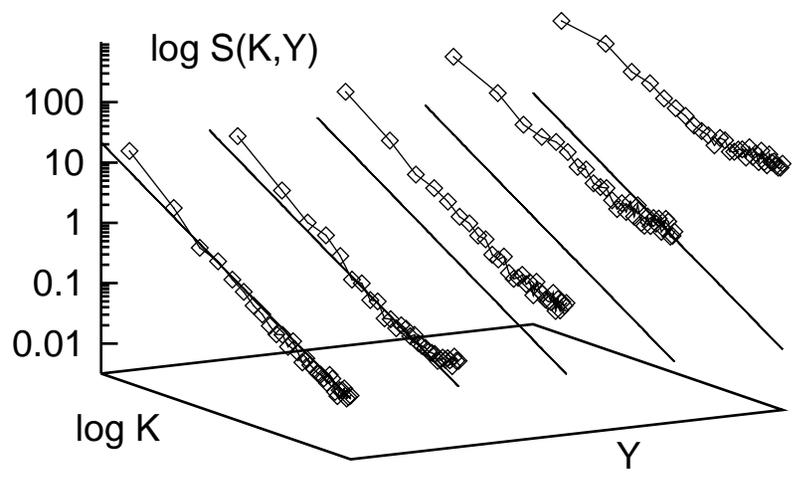
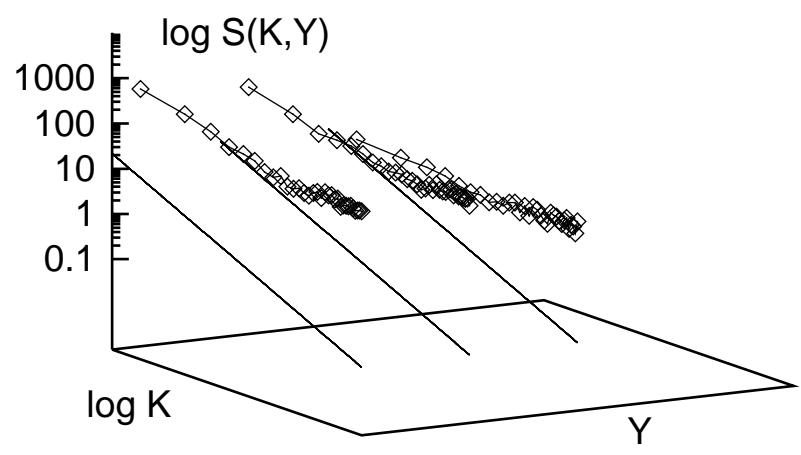
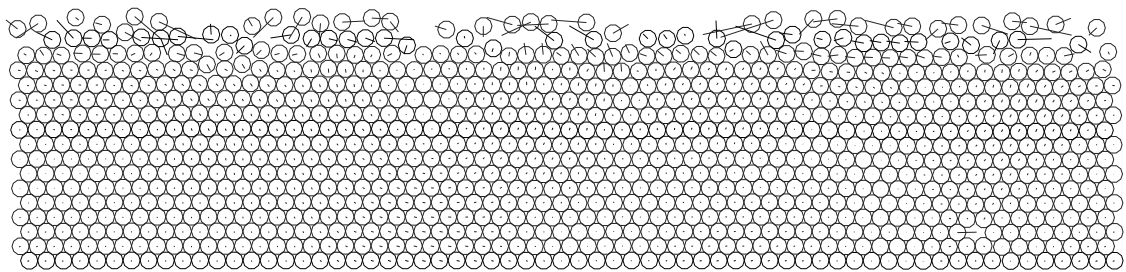
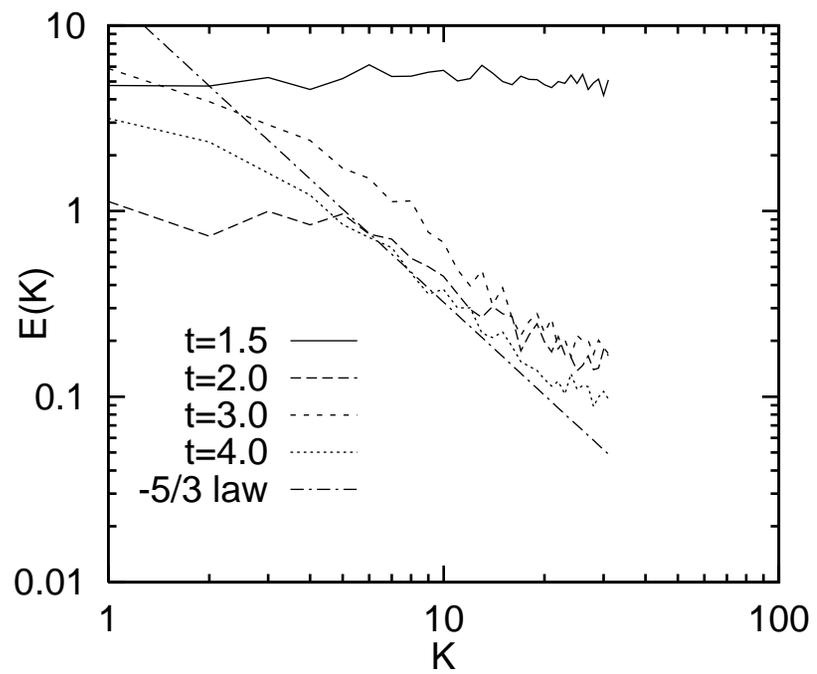
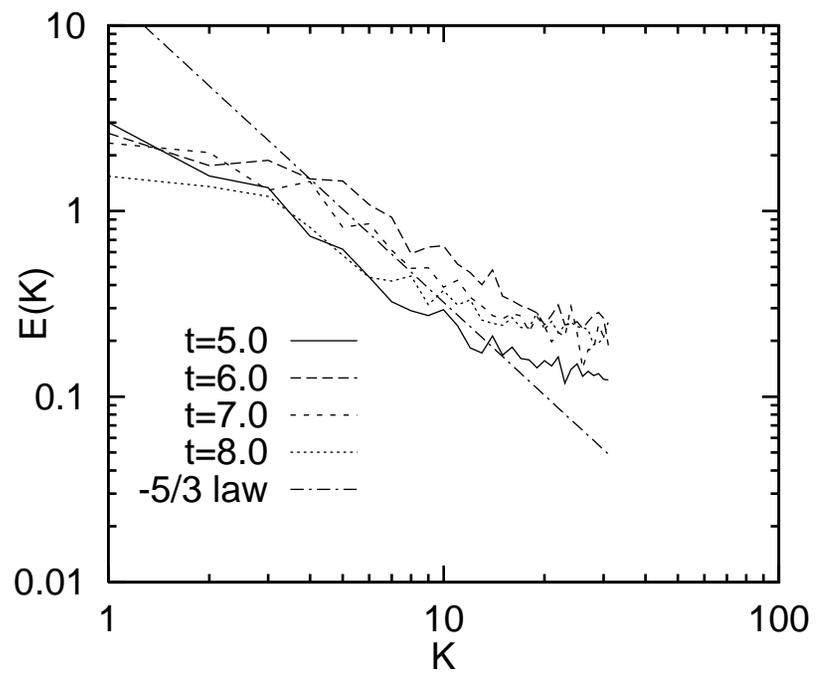
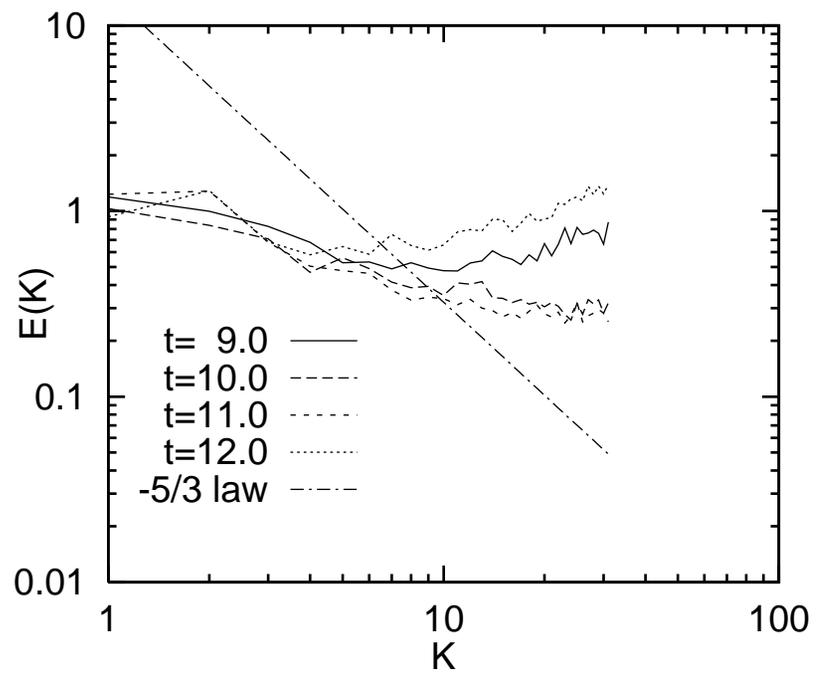
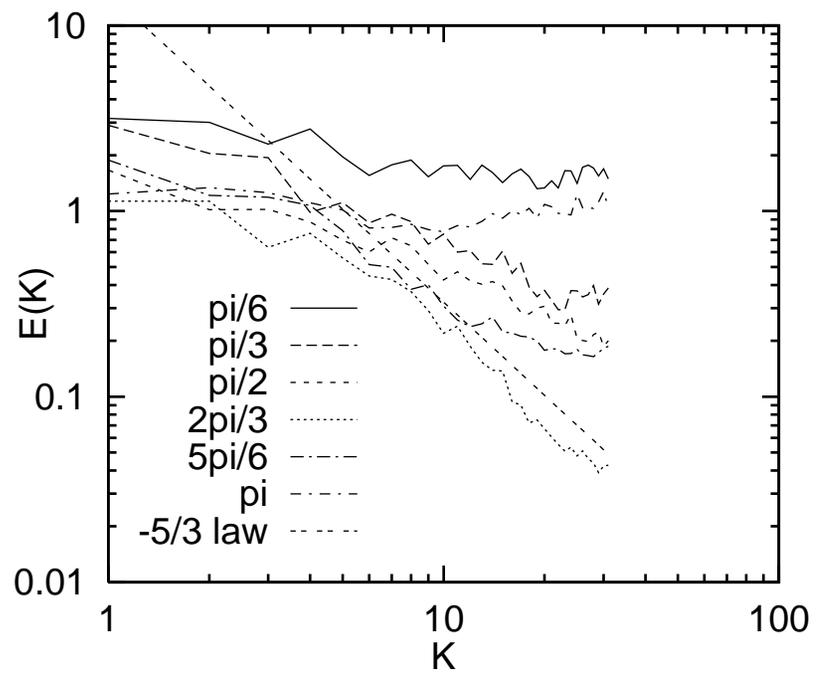
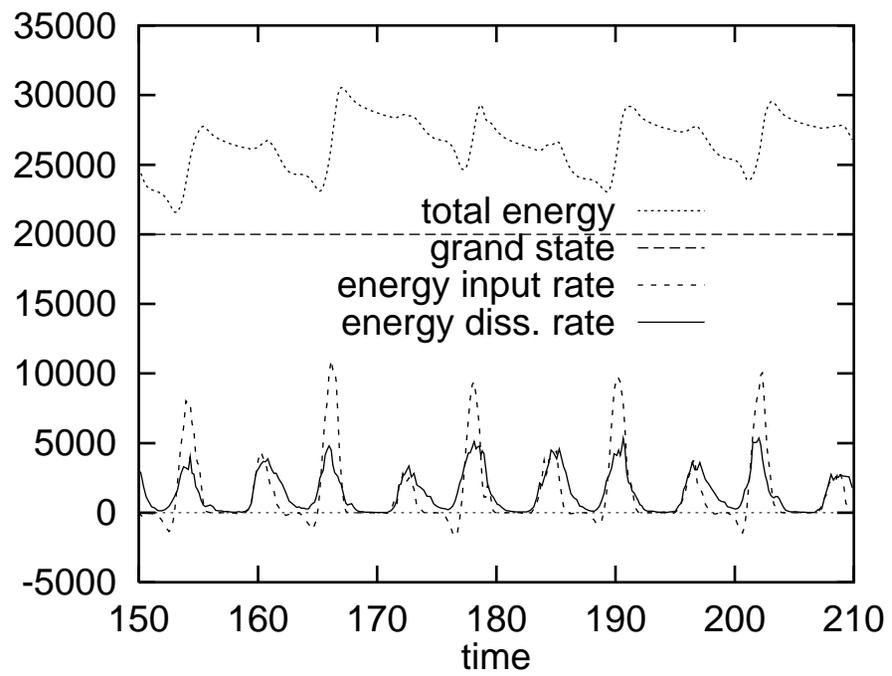
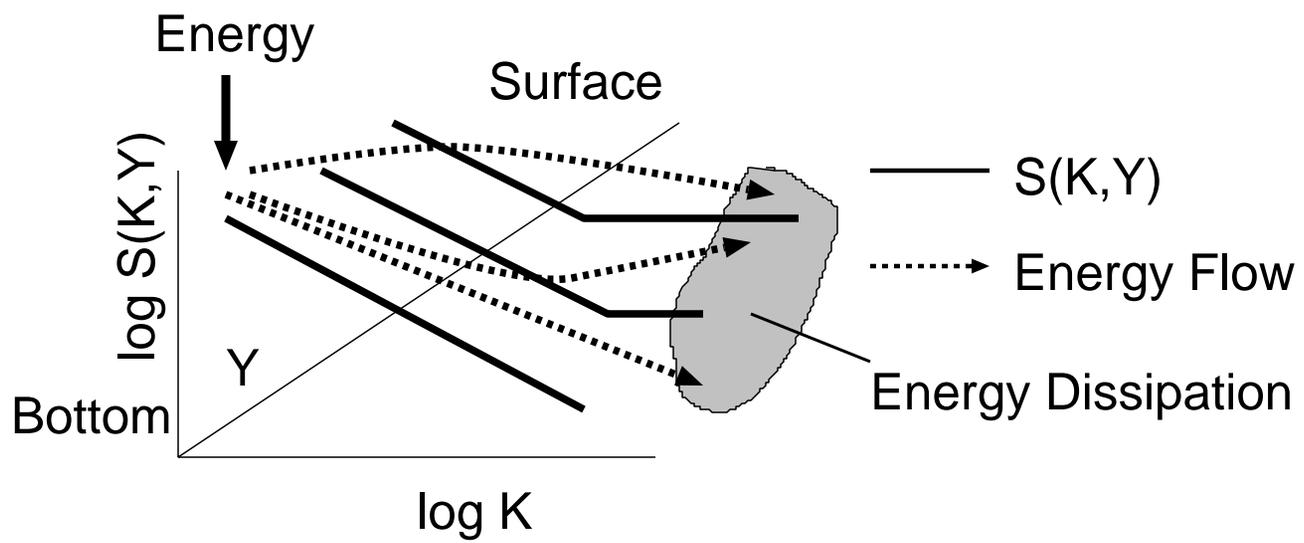
\end{document}